\documentclass[a4paper]{article}

\usepackage{INTERSPEECH2022}

\usepackage{subcaption}
\usepackage{url}
\usepackage{cite}
\title{Decoupled Pronunciation and Prosody Modeling in Meta-Learning-Based Multilingual Speech Synthesis}
\name{Yukun Peng, Zhenhua Ling\thanks{This work was supported in part by the National Key R\&D Program of China under Grant 2019YFF0303001, and in part by the National Nature Science Foundation of China under Grant 61871358.}}
\address{
National Engineering Research Center of Speech and Language Information Processing, \\
University of Science and Technology of China, Hefei, China}
\email{pyk@mail.ustc.edu.cn, zhling@ustc.edu.cn}

\begin{document}

\maketitle
\begin{abstract}
This paper presents a method of decoupled pronunciation and prosody modeling to improve the performance of meta-learning-based multilingual speech synthesis. The baseline meta-learning synthesis method adopts a single text encoder with a parameter generator conditioned on language embeddings and a single decoder to predict mel-spectrograms for all languages. In contrast, our proposed method designs a two-stream model structure that contains two encoders and two decoders for pronunciation and prosody modeling, respectively, considering that the pronunciation knowledge and the prosody knowledge should be shared in different ways among languages. In our experiments, our proposed method effectively improved the intelligibility and naturalness of multilingual speech synthesis comparing with the baseline meta-learning synthesis method.
\end{abstract}

\section{Introduction}

In recent years, neural text-to-speech (TTS) synthesis has achieved remarkable progress \cite{wang2017tacotron, shen2018natural}, and the naturalness of synthetic speech has been improved significantly. 
The acoustic model is a key component in neural TTS systems, which predicts acoustic features from input texts.
One challenge of building acoustic models for multilingual TTS is the difficulty of constructing large-scale speech corpora for all languages, especially for a lot of minor languages in the world.
Therefore, instead of training separate acoustic models for different languages, some studies resorted to building a unified acoustic model using a multilingual training dataset and sharing some model parameters among different languages \cite{zhang2019learning, sitaram2016experiments, xue2019building, li2016multi,nachmani2019unsupervised}.

%With the increasingly close international communication, the traditional single lingual TTS model has been unable to meet our needs. However, extending such models to support multiple, unrelated languages is nontrivial. Some studies \cite{nachmani2019unsupervised, li2016multi} use multiple sets of single TTS engines to process different languages separately. These methods have no obvious advantage over monolingual TTS models, especially when the language data is unbalanced or small. To improve this problem, Some researchers \cite{zhang2019learning, sitaram2016experiments, xue2019building} train multiple language data with a single model. This approach of multilingual hybrid training on one model enables the model to learn knowledge in multiple languages better than training each language individually. However, complete sharing of parameters makes it difficult to learn language differences.

%Considering the differences in multilingual text representations, 
Considering the difficulties of sharing knowledge among the text encoders for different languages, Nekvinda et al. \cite{nekvinda2020one} proposed to replace the original LSTM-based text encoder of Tacotron2 with a meta-learning encoder.
% that adopted the idea of contextual parameter generator network . 
The parameters in the meta-learning encoder were not trained separately for different languages, but were estimated by a parameter generator \cite{platanios2018contextual} conditioned on language embeddings, therefore it can better capture 
%The meta-learn encoder can adjust the parameters of the encoder to learn 
the commonality among languages. 
This method achieved better performance than building a single model for each language and building a unified model but with separate encoders for different languages \cite{nekvinda2020one}.
%This meta-learning approach has good performance in dealing with textual differences.
%However, the meta-learning TTS method only uses the same parameter generator network encoder to learn all the features of the synthesised speech, such as language prosody features and pronunciation features. Language differences in these features are ignored.
One issue with this meta-learning-based multilingual TTS method is that only a single parameter generator was used in the encoder. Thus, it ignored that the pronunciation knowledge and the prosody knowledge should be shared in different ways among languages.

Pronunciation and prosody are two important characteristics of languages. 
The pronunciation differences among languages can be described by their different but overlapped phoneme sets. 
Several studies \cite{xue2019building, chen2019cross, cao2019end} have shown that replacing characters with phonemes as input can significantly improve the pronunciation accuracy of multilingual TTS. 
%On the other hand, different languages extending prosody learning models to support multiple languages may encounter obstacles, such as differences in language prosody features. In speech synthesis, one prominent aspect of foreign accents is a deviant prosody realization \cite{sundstrom1998automatic}. 
On the other hand, it is necessary to consider the prosody properties of different languages when building multilingual TTS systems.
%In , they propose to 
For example, adding tone embeddings was proposed to improve the naturalness of synthesizing tones in Chinese \cite{liu2020tone}. 
Predicting a binary fundamental frequency profile for each phoneme was employed to enhance Japanese synthesis performance in a multilingual model \cite{zhan2021improve}. %However, these optimizations for a single language in multiple languages cannot be extended to other languages trained together. 
Considering that some languages may have similar phoneme sets, while others may have similar prosodic properties, it is reasonable to share the pronunciation knowledge and the prosody knowledge among languages separately in a unified multilingual TTS model.

%In this paper, we investigate different prosody features, such as stress and tone, in multiple languages, while also focusing on commonalities in language pronunciation. 
Therefore, this paper proposes to decouple the pronunciation and prosody modeling in meta-learning-based multilingual speech synthesis.
%to handle pronunciation and prosody information independently and flexibly. 
First, unlike the baseline meta-learning TTS method \cite{nekvinda2020one}, which used character input, our input sequence includes 
%International Phonetic Alphabet (IPA) \cite{international1999handbook} 
International Phonetic Alphabet (IPA) symbols with word boundaries and prosody labels %to maintain the commonality of language pronunciation while 
to introduce explicitly pronunciation-related and prosody-related descriptions. 
Second, a two-stream meta-learning-based model structure is designed, which 
has two encoders and two decoders for pronunciation and prosody modeling, respectively, and a shared attention module. 
Third, instead of using Mel-spectrograms, 
spectral features (i.e., Mel-cepstra) and excitation features (i.e., energy, fundamental frequency and voiced/unvoiced flag) are used as the prediction targets of the pronunciation stream and the prosody stream, respectively. Experimental results show that our proposed method significantly improved the intelligibility and naturalness of the baseline meta-learning multilingual TTS method \cite{nekvinda2020one}.

\section{Proposed method}

In this section, we introduce the input and output representations, the architecture and the training strategy of our proposed acoustic model. 
%propose a two-stream prosody and pronunciation decoupling multilingual text-to-speech model. The proposed method by extending the meta-learning idea from meta-learning TTS \cite{nekvinda2020one}. 
For vocoding, a HiFi-GAN \cite{kong2020hifi} vocoder is used in our implementation. 

\subsection{Input and output representations}

In order to explicitly reflect the pronunciation similarity among different languages, we use IPA phonetic symbols as the basis of model input. Texts are converted to IPA symbol sequences by the Phonemizer tool\footnote{\url{https://github.com/bootphon/phonemizer}}. 
To introduce prosody descriptions, a specific token is inserted at every word boundary in phoneme sequences. 
Besides, a prosody label is assigned to each phoneme to describe the tone or stress character of the phoneme. 
The prosody label is a one-hot vector with \(M+N\) dimensions, where \(M\) corresponds to the number of tones of the tonal languages, and \(N\) corresponds to the number of stress categories of the non-tonal languages.
 
The previous meta-learning-based multilingual synthesis model \cite{nekvinda2020one} adopted Mel-spectrograms as model output, which mixes all pronunciation-related and prosody-related information. 
In our proposed method, different acoustic features are utilized to represent these two types of information. 
Specifically, spectral features, i.e., Mel-cepstra, are used as the output of the pronunciation stream in our model, while excitation features, i.e., energy, logarithmic F0 (logF0) and voiced/unvoiced (V/UV) flag are used as the output of the prosody stream in our model.
These acoustic features are extracted by the STRAIGHT vocoder \cite{kawahara1999restructuring}.
All features are normalized to zero mean and unit variance except the V/UV flag before acoustic modeling.

\begin{figure}[t]
\centering
\includegraphics[width=\linewidth,]{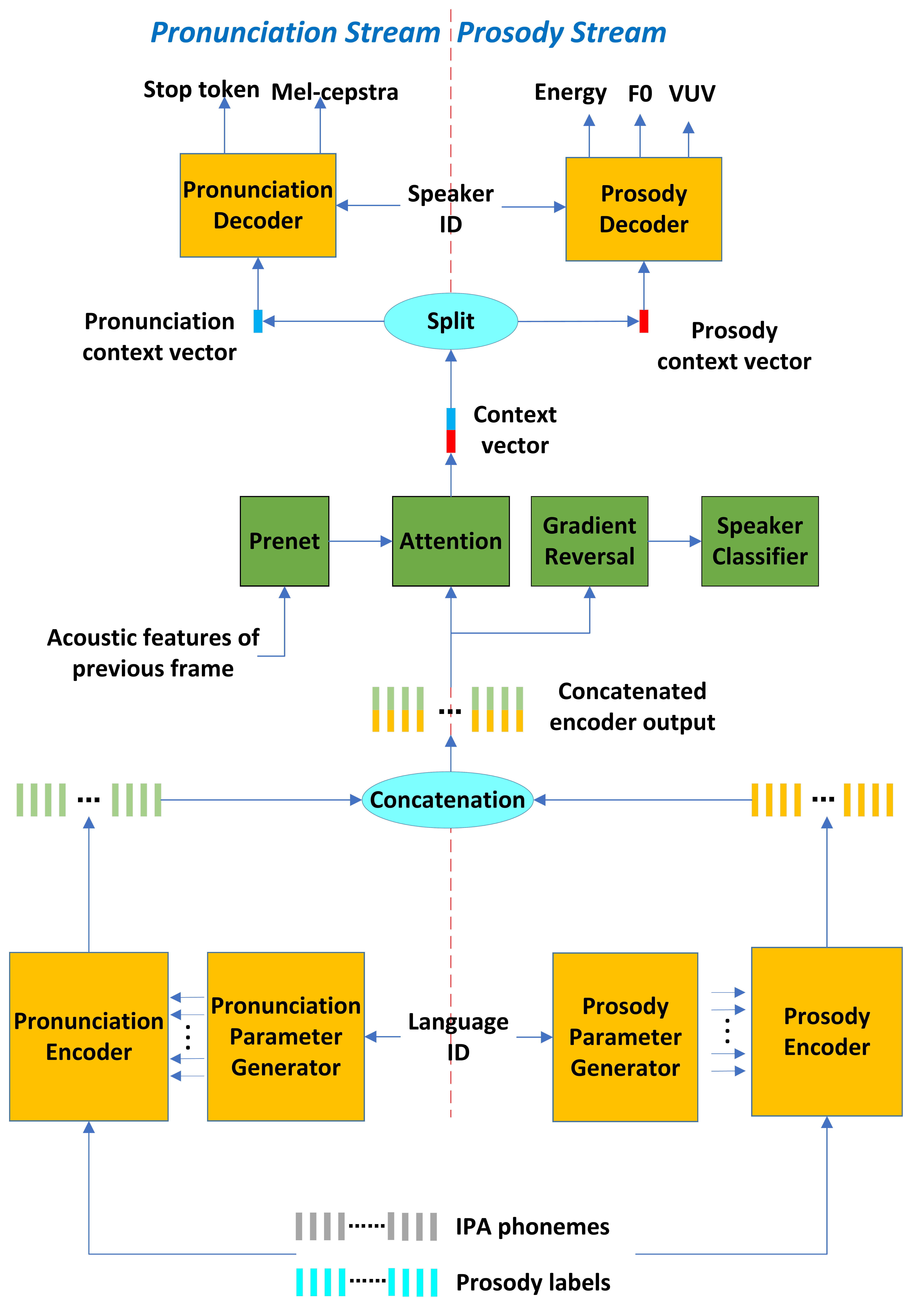}
\vspace{-6mm}
\caption{The architecture of the proposed model. }
\label{fig:model_figure}
\vspace{-6mm}
\end{figure}

\subsection{Model architecture}

The architecture of our proposed model %designed to decouple prosody and pronunciation modeling 
is shown in Fig.~\ref{fig:model_figure}.
%Compared with the original meta-learning TTS, our main improvement is that we design 
It follows the attention-based sequence-to-sequence (seq2seq) framework for acoustic modeling and adopts Tacotron2 \cite{shen2018natural} as its backbone. 
%A pronunciation stream and a prosody stream are designed to achieve pronunciation and prosody modeling among multiple languages, respectively. 
As shown in Fig.~\ref{fig:model_figure},
it contains a pronunciation stream and a prosody stream.
These two streams have separate text encoders with separate parameter generators conditioned on language embeddings, and separate decoders for predicting different acoustic features. 
Two streams share an attention module to maintain the synchrony between the two streams.
 
In each encoder, a lookup table is employed to convert IPA phonemes and prosody labels into IPA embedding vectors and prosody embedding vectors, respectively. For each phoneme, the IPA embedding vector and the prosody embedding vector are concatenated and are then sent into the encoder. 
%the final encoder input vector. 
The pronunciation encoder and the prosody encoder contain two separate embedding tables,
which allows the model to select necessary information from the input in a stream-dependent manner. %we model the prosody hidden representation and pronunciation hidden representation, from IPA and prosody embedding, via two meta-learning encoders. 
Each encoder includes two 1-dimensional-convolutional (1D-Conv) layers and twelve highway 1D-Conv layers. The stride and size of the convolution kernel of each layer refer to DCTSS \cite{tachibana2018efficiently}. 
Each encoder relies on a parameter generator %layer of convolutional network uses fully connected layers as a generator 
to obtain the weights and biases of its network. 
Following previous study \cite{nekvinda2020one}, each parameter generator takes language embeddings as input.
%, and generates specific parameters for the encoder according to the language. 
Let \({\textbf{X}_a} \in \textbf{R}^{{D_a}*L}\) and \({\textbf{X}_p} \in \textbf{R}^{{D_p}*L}\) denote the outputs of the pronunciation encoder and the prosody encoder, where \({{D_a}}\) and \({{D_p}}\) are the output dimensions of the two encoders and %pronunciation encoder, \({{D_p}}\) is the dimensions of prosody encoder and 
\(L\) is phoneme sequence length. 
%To obtain same alignment weights, 
Then, \(\textbf{X}_a\) and \(\textbf{X}_p\) are concatenated to get \({\textbf{X}} \in \textbf{R}^{{(D_a+D_p)}*L}\) for attention calculation.

%For all languages, we share the same attention model. 
The attention module is shared by both streams to keep inter-stream synchrony.
Following previous work \cite{zhan2021improve}, location sensitive attention mechanism \cite{chorowski2015attention} is adopted here to align text encoder outputs to acoustic feature sequences. The predicted mel-cepstra, logF0 and energy at the previous frame are passed to the prenet. %as autoregressive acoustic features of the previous frame. 
Following Tacotron2 \cite{shen2018natural}, a long short-term memory (LSTM) layer is applied to obtain query vectors from the prenet output and the context vector of previous frame. % attention context. 
The concatenated encoder outputs act as keys and values of attention calculation.
%Then, the attention module predicts attention weights and context vectors from the output of the LSTM and the concatenated encoder output. 
%In order for the prosody stream to learn the prosody of phonemes instead of doing the same work as pronunciation part, 
Then, the derived context vector at each frame is further split into two parts according to \({{D_a}}\) and \({{D_p}}\), 
%by encoders' dimensions(see figure. ~\ref{fig:model_figure}), 
which are sent into the decoders of two streams, respectively. 

Each decoder contains an LSTM structure. In the prosody decoder, a lookup table converts the speaker ID into a speaker embedding vector, concatenated with context vectors as the input to LSTM. The LSTM output is projected through two separate linear transformations to predict energy and logF0, respectively. Meanwhile, the output is also projected into a scalar by a linear layer with sigmoid activation to predict a V/UV flag. The pronunciation decoder uses the same structure to predict a mel-cepstrum vector and a stop flag, respectively. 

Following the previous work \cite{zhan2021improve}, an adversarial speaker classifier \cite{jia2018transfer} with a gradient reversal layer is applied to the concatenated encoder output. It follows the principle of domain adversarial training \cite{ganin2016domain} to remove the residual speaker information in the encoder output.

\subsection{Model training}
All model parameters are estimated simultaneously with a multilingual training corpus.
The training loss consists of two parts. One part is the loss of reconstructing acoustic features, i.e., \(Los{s_{Rec}}\). Mean squared error (MSE) loss function is adopted for mel-cepstra, energy and LogF0 prediction, while binary cross entropy (BCE) loss function is used for V/UV flag and stop flag prediction. % The loss function of the reconstruction loss of acoustic features can be expressed as:
These loss functions are added to get \(Los{s_{Rec}}\).
% %
% \begin{equation}
% Los{s_{Rec}} = MS{E_{mcep}} + MS{E_{F0}} + MS{E_{en}} + BC{E_{uv}} + BC{E_{stop}}
% \end{equation}
% %
The other part is the speaker classifier loss, i.e., \( Los{s_{Spk }}\), following Zhang et al.'s work \cite{zhang2019learning}.
% % 
% \begin{equation}
% Los{s_{Speaker }}({\psi _S};{\textbf{t}_i}) = \sum\limits_i^N {\Log p} ({\textbf{s}_i}|{\textbf{t}_i})
% \end{equation}
% % 
% where \({\textbf{s}_i}\) is the speaker label, \({\psi _S}\) are the parameters for speaker classifier and \({\textbf{t}_i}\) is the text encoding. 
The final loss function of our proposed model can be written as
\vspace{-0.6\baselineskip}
\begin{equation}
Los{s_{Total}} = Los{s_{Rec}} - \lambda Los{s_{Spk }},
\vspace{-0.6\baselineskip}
\end{equation}
where $\lambda$ is a weight tuned manually in our implementation.
% 

% Besides, in training step, To ensure language balance, our batch size is set to be divisible by the number of languages \(L\), and all \((i + nL)\) in a batch are the same language, where \(i < L\), \(n < L/B\). 
% In addition, the real mel-cepstra, LogF0 and energy will be used as the teacher forcing input of the prenet. We did not use V/UV flags as prenet input, because introducing V/UV flags into prenet would make the model predict the V/UV flags result of the next frame directly from the V/UV flags of the previous frame instead of judging from the representation. 

%Aliquam quis orci consectetur nulla luctus ullamcorper. Suspendisse finibus luctus erat a dapibus. 

\section{Experiments}

\subsection{Dataset}

Similar to previous study \cite{nekvinda2020one}, our experiments used a subset of the multilingual single speaker dataset CSS10 \cite{park2019css10} and selected clear speakers from the multilingual multi-speaker dataset Common Voice \cite{ardila2020common} to enhance CSS10. There are 10 languages in the original CSS10 dataset, and we used five of them in our experiments, including Mandarin (ZH), German (DE), French (FR), Dutch (NL) and Russian (RU). %To reduce the impact of batch imbalance, 
We removed too long and too short sentences in the dataset by setting the maximum and minimum sentence durations to 10s and 1s. Table~\ref{tab:dataset} shows the data amount used for experiment. Then, the data of each language was split into a training set, a development set and a test set with a ratio of 8:1:1. All audios were sampled at 22.05 kHz. Our Hi-FiGAN vocoder was trained on the training sets of all languages. 
\begin{table}[t]
\centering
\caption{The data amount of the CSS/Common Voice datasets used in our experiments.}
\vspace{-3mm}
\resizebox{\linewidth}{!}{%
\small
\begin{tabular}{cccccc}
\toprule[1.5pt]
\multicolumn{1}{c|}{\textbf{Language}} & \multicolumn{1}{c|}{\textbf{ZH}} & \multicolumn{1}{c|}{\textbf{DE}} & \multicolumn{1}{c|}{\textbf{FR}} & \multicolumn{1}{c|}{\textbf{NL}} & \multicolumn{1}{c}{\textbf{RU}} \\ \hline
\multicolumn{1}{c|}{\textbf{hours}} & \multicolumn{1}{c|}{6.5/1.0} & \multicolumn{1}{c|}{16.1/4.8} & \multicolumn{1}{c|}{19.2/3.0} & \multicolumn{1}{c|}{14.1/1.3} & \multicolumn{1}{c}{21.4/3.4} \\ \hline
\multicolumn{1}{c|}{\textbf{speakers}} & \multicolumn{1}{c|}{1/6} & \multicolumn{1}{c|}{1/39} & \multicolumn{1}{c|}{1/22} & \multicolumn{1}{c|}{1/11} & \multicolumn{1}{c}{1/8} \\ 
\bottomrule[1.5pt]
\end{tabular}
\label{tab:dataset}%
}
\vspace{-6mm}
\end{table}

\subsection{Model implementation}

STRAIGHT \cite{kawahara1999restructuring} was applied to extract acoustic features, which included 40-dimensional mel-cepstra, 
an energy, an F0 and a V/UV flag for each frame. 
The frame length was 25ms, and the frame shift was 10ms.
%computed from 25ms windows shifted by 10ms. 
These total 43-dimensional features were used as the input of our HiFi-GAN vocoder. 

For the prosody labels, we had $M$=5 according to the five tones in Mandarin and $N$=3 according to the stress categories of non-tonal languages.
Here, the stress categories included primary stressed vowel, secondary stressed vowel and non-stressed phoneme.
% voiced consonant and unvoiced consonant.
The dimensions of IPA embeddings and prosody label embeddings were 512 and 16.
%The IPA sequence and prosody labels sequence are passed through lookup table output 512-dim IPA embedding vectors and 16-dim prosody embedding vectors. 
The pronunciation encoder model had 14 1D-Conv layers with $D_a=256$ in each layer, like DCTTS \cite{tachibana2018efficiently}, while the prosody encoder model set $D_p=128$ in each layer due to the low dimensionality of prosody features. The hidden
unit numbers in the pronunciation decoder and the prosody decoder were 1024 and 256. We set the batch size to 50 and considered the language balance when composing each batch. In the prosody encoder and the prosody decoder, half of the initial learning rate was used for parameter updating to reduce overfitting. The learning rate of the remaining model parameters was initialized to \(10^{-3}\). The Adam optimizer was adopted %with a learning rate decay strategy, 
and the learning rate decayed to half every 15000 steps. The weight \(\lambda \) in Eq. (1) was set to 0.05. 
%Furthermore, since the depth of the features is low enough, we do not need postnet from Tacotron2 model. 

\begin{table}[t!]
\centering
\caption{Objective evaluation results of different models.}
\vspace{-3mm}
\label{tab:Objective}
% \resizebox{\linewidth}{!}{%

\begin{subtable}[t]{\linewidth}
\caption{Mel-cepstrum distortion (MCD) (dB)}
\vspace{-3mm}
\resizebox{\linewidth}{!}{%
\begin{tabular}{ccccc}
\midrule[1.5pt]
\multicolumn{1}{c|}{\textbf{Language}} & \multicolumn{1}{c|}{\textbf{Tacotron2}} & \multicolumn{1}{c|}{\textbf{Meta-char}} & \multicolumn{1}{c|}{\textbf{Meta-IPA}} & \multicolumn{1}{c}{\textbf{Proposed}} \\ \hline
\multicolumn{1}{c|}{\textbf{ZH}}    & \multicolumn{1}{c|}{2.770}       & \multicolumn{1}{c|}{2.563}       & \multicolumn{1}{c|}{2.460}       & \multicolumn{1}{c}{\textbf{2.445}}  \\ %\hline
\multicolumn{1}{c|}{\textbf{DE}}    & \multicolumn{1}{c|}{3.240}       & \multicolumn{1}{c|}{3.111}       & \multicolumn{1}{c|}{2.992}       & \multicolumn{1}{c}{\textbf{2.788}}  \\ %\hline
\multicolumn{1}{c|}{\textbf{FR}}    & \multicolumn{1}{c|}{3.184}       & \multicolumn{1}{c|}{2.984}       & \multicolumn{1}{c|}{2.814}       & \multicolumn{1}{c}{\textbf{2.784}}  \\ %\hline
\multicolumn{1}{c|}{\textbf{NL}}    & \multicolumn{1}{c|}{3.074}       & \multicolumn{1}{c|}{2.953}       & \multicolumn{1}{c|}{2.905}       & \multicolumn{1}{c}{\textbf{2.872}}  \\ %\hline
\multicolumn{1}{c|}{\textbf{RU}}    & \multicolumn{1}{c|}{4.074}       & \multicolumn{1}{c|}{3.719}       & \multicolumn{1}{c|}{\textbf{3.449}}  & \multicolumn{1}{c}{3.509}      \\ 
\bottomrule[1.5pt]
\\
\end{tabular}}
\end{subtable}
\begin{subtable}[t]{\linewidth} \vspace{-2mm}
\caption{Root mean square error of F0 (F0-RMSE) (Hz)}
\vspace{-3mm}
\resizebox{\linewidth}{!}{%
\begin{tabular}{ccccc}
\midrule[1.5pt]
% \multicolumn{5}{c}{F0-RMSE} \\ \hline
\multicolumn{1}{c|}{\textbf{Language}} & \multicolumn{1}{c|}{\textbf{Tacotron2}} & \multicolumn{1}{c|}{\textbf{Meta-char}} & \multicolumn{1}{c|}{\textbf{Meta-IPA}} & \multicolumn{1}{c}{\textbf{Proposed}} \\ \hline
\multicolumn{1}{c|}{\textbf{ZH}} & \multicolumn{1}{c|}{44.591}& \multicolumn{1}{c|}{37.876} & \multicolumn{1}{c|}{35.006}& \multicolumn{1}{c}{\textbf{33.268}}\\ %\hline
\multicolumn{1}{c|}{\textbf{DE}} & \multicolumn{1}{c|}{44.880}& \multicolumn{1}{c|}{43.776} & \multicolumn{1}{c|}{38.722}& \multicolumn{1}{c}{\textbf{37.220}}\\ %\hline
\multicolumn{1}{c|}{\textbf{FR}} & \multicolumn{1}{c|}{23.418}& \multicolumn{1}{c|}{23.686} & \multicolumn{1}{c|}{20.260}& \multicolumn{1}{c}{\textbf{18.677}}\\ %\hline
\multicolumn{1}{c|}{\textbf{NL}} & \multicolumn{1}{c|}{38.374}& \multicolumn{1}{c|}{38.267} & \multicolumn{1}{c|}{32.385}& \multicolumn{1}{c}{\textbf{31.743}}\\ %\hline
\multicolumn{1}{c|}{\textbf{RU}} & \multicolumn{1}{c|}{51.631}& \multicolumn{1}{c|}{48.365} & \multicolumn{1}{c|}{43.850}& \multicolumn{1}{c}{\textbf{41.624}}\\ 
% \multicolumn{5}{c}{}\\ %\hline
\bottomrule[1.5pt]
\\
\end{tabular}}
\end{subtable}

\begin{subtable}[t]{\linewidth} \vspace{-2mm}
\caption{Pearson correlation coefficient of F0 (F0-CORR)}
\vspace{-2mm}
\resizebox{\linewidth}{!}{%
\begin{tabular}{ccccc}
\toprule[1.5pt]
% \multicolumn{5}{c}{F0-CORR} \\ \hline
\multicolumn{1}{c|}{\textbf{Language}} & \multicolumn{1}{c|}{\textbf{Tacotron2}} & \multicolumn{1}{c|}{\textbf{Meta-char}} & \multicolumn{1}{c|}{\textbf{Meta-IPA}} & \multicolumn{1}{c}{\textbf{Proposed}} \\ \hline
\multicolumn{1}{c|}{\textbf{ZH}} & \multicolumn{1}{c|}{0.424} & \multicolumn{1}{c|}{0.604}& \multicolumn{1}{c|}{0.654} & \multicolumn{1}{c}{\textbf{0.683}} \\ %\hline
\multicolumn{1}{c|}{\textbf{DE}} & \multicolumn{1}{c|}{0.275} & \multicolumn{1}{c|}{0.357}& \multicolumn{1}{c|}{0.479} & \multicolumn{1}{c}{\textbf{0.522}} \\ %\hline
\multicolumn{1}{c|}{\textbf{FR}} & \multicolumn{1}{c|}{0.309} & \multicolumn{1}{c|}{0.370}& \multicolumn{1}{c|}{0.523} & \multicolumn{1}{c}{\textbf{0.566}} \\ %\hline
\multicolumn{1}{c|}{\textbf{NL}} & \multicolumn{1}{c|}{0.284} & \multicolumn{1}{c|}{0.349}& \multicolumn{1}{c|}{0.479} & \multicolumn{1}{c}{\textbf{0.514}} \\ %\hline
\multicolumn{1}{c|}{\textbf{RU}} & \multicolumn{1}{c|}{0.094} & \multicolumn{1}{c|}{0.226}& \multicolumn{1}{c|}{0.354} & \multicolumn{1}{c}{\textbf{0.430}}\\ 
\bottomrule[1.5pt]
\\
\end{tabular}}
\end{subtable}

\begin{subtable}[t]{\linewidth} \vspace{-2mm}
\caption{Root mean square error of energy (EN-RMSE) (Hz)}
\vspace{-2mm}
\resizebox{\linewidth}{!}{%
\begin{tabular}{ccccc}
% \multicolumn{5}{c}{EN-RMSE} \\ \hline
\toprule[1.5pt]
\multicolumn{1}{c|}{\textbf{Language}} & \multicolumn{1}{c|}{\textbf{Tacotron2}} & \multicolumn{1}{c|}{\textbf{Meta-char}} & \multicolumn{1}{c|}{\textbf{Meta-IPA}} & \multicolumn{1}{c}{\textbf{Proposed}} \\ \hline
\multicolumn{1}{c|}{\textbf{ZH}} & \multicolumn{1}{c|}{0.234} & \multicolumn{1}{c|}{0.201}& \multicolumn{1}{c|}{0.174} & \multicolumn{1}{c}{\textbf{0.166}} \\ %\hline
\multicolumn{1}{c|}{\textbf{DE}} & \multicolumn{1}{c|}{0.261} & \multicolumn{1}{c|}{0.254}& \multicolumn{1}{c|}{0.238} & \multicolumn{1}{c}{\textbf{0.188}} \\ %\hline
\multicolumn{1}{c|}{\textbf{FR}} & \multicolumn{1}{c|}{0.260} & \multicolumn{1}{c|}{0.250}& \multicolumn{1}{c|}{0.221} & \multicolumn{1}{c}{\textbf{0.218}} \\ %\hline
\multicolumn{1}{c|}{\textbf{NL}} & \multicolumn{1}{c|}{0.204} & \multicolumn{1}{c|}{0.202}& \multicolumn{1}{c|}{\textbf{0.180}}& \multicolumn{1}{c}{0.195}\\ %\hline
\multicolumn{1}{c|}{\textbf{RU}} & \multicolumn{1}{c|}{0.362} & \multicolumn{1}{c|}{0.324}& \multicolumn{1}{c|}{0.280} & \multicolumn{1}{c}{\textbf{0.270}} \\ 
% \multicolumn{5}{c}{}\\ %\hline
\bottomrule[1.5pt]
\\
\end{tabular}}
\end{subtable}

\begin{subtable}[t]{\linewidth} \vspace{-2mm}
\caption{V/UV flag error rate (V/UV-ERR) ($\%$)}
\vspace{-2mm}
\resizebox{\linewidth}{!}{%
\begin{tabular}{ccccc}
% \multicolumn{5}{c}{UV-ERR} \\ \hline
\toprule[1.5pt]
\multicolumn{1}{c|}{\textbf{Language}} & \multicolumn{1}{c|}{\textbf{Tacotron2}} & \multicolumn{1}{c|}{\textbf{Meta-char}} & \multicolumn{1}{c|}{\textbf{Meta-IPA}} & \multicolumn{1}{c}{\textbf{Proposed}} \\ \hline
\multicolumn{1}{c|}{\textbf{ZH}} & \multicolumn{1}{c|}{10.665}& \multicolumn{1}{c|}{8.854}& \multicolumn{1}{c|}{8.351} & \multicolumn{1}{c}{\textbf{7.696}} \\ %\hline
\multicolumn{1}{c|}{\textbf{DE}} & \multicolumn{1}{c|}{10.248}& \multicolumn{1}{c|}{10.589} & \multicolumn{1}{c|}{9.093} & \multicolumn{1}{c}{\textbf{7.139}} \\ %\hline
\multicolumn{1}{c|}{\textbf{FR}} & \multicolumn{1}{c|}{14.770}& \multicolumn{1}{c|}{14.313} & \multicolumn{1}{c|}{10.316}& \multicolumn{1}{c}{\textbf{8.884}} \\ %\hline
\multicolumn{1}{c|}{\textbf{NL}} & \multicolumn{1}{c|}{17.108}& \multicolumn{1}{c|}{16.223} & \multicolumn{1}{c|}{16.785}& \multicolumn{1}{c}{\textbf{16.161}}\\ %\hline
\multicolumn{1}{c|}{\textbf{RU}} & \multicolumn{1}{c|}{19.911}& \multicolumn{1}{c|}{17.294} & \multicolumn{1}{c|}{14.234}& \multicolumn{1}{c}{\textbf{13.304}}\\ 
\bottomrule[1.5pt]
\end{tabular}}
\end{subtable}
\vspace{-5mm}
\end{table}

\subsection{Baseline models}

To verify the effectiveness of our proposed method, three baseline models were built for comparison. For a fair comparison, the output acoustic features of baseline models were the same as the 43-dimensional ones used in our proposed model and were optimized by the same loss function.
%pronunciation and prosody features. 
Preliminary experimental results showed that using 43-dimensional acoustic features or 80-dimensional Mel-spectrograms in this model led to similar performance of synthetic speech.

\noindent\textbf{Tacotron2}: This model followed the original Tacotron2 architecture \cite{shen2018natural}. To be compatible with multilingual speech synthesis, it had a fully shared encoder with characters and a language ID as input. %, Using , 
Similar to our proposed model, an adversarial speaker classifier was added to remove the speaker information contained in the encoder output, and a speaker embedding was connected to the input state of the LSTM decoder layer. Its hyperparameters %of the attention and decoder modules
 were consistent with the ones in our proposed model.

\noindent\textbf{Meta-char}: This model was built following the baseline meta-learning-based multilingual TTS method %Meta-char is a meta-learning TTS model, detailed setting reference 
\cite{nekvinda2020one}. including meta-learning encoder and character input. 
%Compared with Mel-spectrogram features, and there is no significant difference in the results between the two, proving that this replacement is feasible.

\noindent\textbf{Meta-IPA}: %Meta-IPA based on meta-learning TTS model. However, we choose IPA and prosody labels as input features to compare their contribution to the model. At the same time, considering that our model has a larger number of parameters, based on Mate-IPA, we did a control experiment of doubling the parameters. The results showed that these changes did not significantly affect the pronunciation and prosody of speech. 
This model was the same as Meta-char, and the only difference was that IPA phonemes and prosody labels were used as model input instead of characters, just like our proposed model.
The difference between Meta-IPA and our proposed model was that the two-stream modeling was adopted in our proposed model. It should be noticed that our proposed model had more model parameters than Meta-char due to the additional encoder and decoder. However, we have conducted some preliminary experiments to confirm that increasing the number of parameters in Meta-IPA accordingly can't improve the performance of this model.

%\section{Experimental Results}

% In our experiments, the Meta-learning TTS model and fully shared Tacotron2 were used as the baseline. %For the baseline models we use IPA instead of characters as input, which we demonstrate in our experiments to reduce skipping and missing words in synthesised speech. 
% Tacotron baseline has a single fully shared encoder like the original Tacotron 2 architecture. Add an adversarial speaker classifier to remove speaker information contained in the encoder output. we concatenate the speaker embedding to the input states of the LSTM decoder layers. We choose characters as the input of the tacotron baseline, and compare the impact of IPA and character input on the meta-learning TTS. To evaluate the prosody of the models, subjective and Objective tests were used. %, i. e. , the mean opinion score (MOS) and preference tests conducted by 10 native speakers, per language. 

\subsection{Objective evaluation}
\label{subsec:objeval}
100 sentences were randomly selected from the test set of each language and were synthesized by our proposed model and baseline models.
An objective evaluation was conducted on the synthetic speech.
The evaluation metrics and results %o compare with real speech. As 
are presented in Table~\ref{tab:Objective}.

From this table, we can see that the two meta-learning-based baselines performed better than Tacotron2, and Meta-IPA outperformed Meta-char. 
Our proposed model achieved the best performance on all metrics and all languages, except that Meta-IPA slightly outperformed our proposed method on the MCD metric of Russian and the EN-RMSE metric of Dutch.
% our method only performed slightly worse than Meta-IPA on Dutch. We think this may be related to the stable energy of the Dutch dataset. A possible reason is the poor quality of the recordings from the Russian dataset \cite{nekvinda2020one}. The results show that our proposed structure has a stronger ability to model and extract prosody and pronunciation attributes.
These results demonstrated the effectiveness of meta-learning-based acoustic model, the strategy of composing IPA phonemes and prosody labels as model input, and the two-stream model structure proposed in this paper on improving the accuracy of acoustic feature prediction.

Furthermore, we evaluated the intelligibility of synthetic utterances by sending them into the speech recognition engine of Google cloud platform\footnote{\url{https://cloud.google.com/speech-to-text}}.
The character error rate (CER) of speech recognition was used as the evaluation metric and the results are shown in Table~\ref{tab:cer}.
From this table, we can see that Tacotron2 had the highest CERs. Meta-IPA performed better than Meta-char, and %the results of Meta-char and Meta-IPA indicate that IPA input and meta structure positively influence pronunciation correctness. 
our proposed model achieved the lowest CERs for %has the highest recognition accuracy in 
all five languages. 
This indicates that in addition to meta-learning and using IPAs, %the effects of IPA and meta-learning, the 
the proposed method of decoupled pronunciation and prosody modeling also benefited 
%cial for the pronunciation module to pay more attention to the pronunciation of IPA phonemes. 
the accurate pronunciation of synthetic speech.

\begin{table}[t]
\centering
\caption{Character error rates (CER) (\%) of different models.}
\vspace{-3mm}
\label{tab:cer}
\resizebox{\linewidth}{!}{%
\begin{tabular}{c|c|c|c|c}
\toprule[1.5pt]
% \hline
\textbf{Language} & \textbf{Tacotron2} & \textbf{Meta-char} & \textbf{Meta-IPA} & \textbf{Proposed} \\
\hline
ZH& 38.4& 28.7& 25.9& \textbf{24.9 } \\
DE& 16.6& 10.1& 7.5 & \textbf{6.4 } \\
FR& 31.3& 21.2& 18.8& \textbf{17.7 } \\
NL& 26.3& 17.7& 16.8& \textbf{15.0 } \\
RU& 39.2& 24.4& 15.6& \textbf{13.3 } \\%\hline
\bottomrule[1.5pt]
\end{tabular}%
}
\vspace{-3mm}
\end{table}%

\begin{table}[t]
	\centering
	\caption{Naturalness mean opinion scores (MOS) of different models with 95$\%$ confidence intervals, where ``GT" means ground truth. }
	\vspace{-3mm}
	\resizebox{\linewidth}{!}{%
		\begin{tabular}{c|c|c|c|c|c}
			\toprule[1.5pt]
			%\hline
			%\textbf{Language} 
			& \textbf{Tacotron2} & \textbf{Meta-char} & \textbf{Meta-IPA} & \textbf{Proposed} & \textbf{GT} \\ \hline
			ZH & 2.25$\pm$0.15& 3.08$\pm$0.15 & 3.25$\pm$0.15& 3.49$\pm$0.14 & 3.69$\pm$0.15 \\
			DE & 2.14$\pm$0.13& 2.50$\pm$0.15& 3.1$\pm$0.13 & 3.30$\pm$0.13& 3.66$\pm$0.13 \\
			FR & 2.53$\pm$0.15& 3.54$\pm$0.14 & 3.24$\pm$0.15& 3.86$\pm$0.12 & 3.95$\pm$0.11 \\
			NL & 3.02$\pm$0.10& 3.36$\pm$0.07 & 3.92$\pm$0.06& 3.96$\pm$0.05 & 4.31$\pm$0.07 \\
			RU & 2.8$\pm$0.17 & 3.52$\pm$0.13 & 3.79$\pm$0.13& 3.96$\pm$0.11 & 4.36$\pm$0.11\\%\hline
			\bottomrule[1.5pt]
		\end{tabular}
		\label{tab:mos}%
	}
\vspace{-5mm}
\end{table}%

\subsection{Subjective evaluation}
A group of subjective listening tests were conducted to evaluate the naturalness mean opinion scores (MOS) of different models. The score range was from 1 (completely unnatural) to 5 (completely natural). %points in 1-point steps, and we use 
20 utterances\footnote{Audio samples can be found at \url{https://pengyuk.github.io/dppmttsdemo}} were used for each model and each language. 
For Mandarin, Russian and Dutch, 11, 7, and 8 native listeners were recruited offline, respectively.
%, 11 offline recruited Mandarin native raters, 7 offline recruited Russian native raters and 8 offline recruited Dutch native raters participated in the scoring. 
For German and French, the tests were conducted by crowdsourcing on Amazon Mechanical Turk\footnote{\url{https://www.mturk.com}}, %including 14 French raters and 10 German raters. 
with 14 and 10 native listeners, respectively. 
The utterances recovered from ground truth acoustic features using the same HiFi-GAN vocoder
were also included for comparison. The results are summarized in Table~\ref{tab:mos}. 
%We also score the ground truth to compare the synthetic speech with the real speech. Ground truth audio is synthesised from real acoustic features. 
We can see that the subjective evaluation results were consistent with the objective ones in Section \ref{subsec:objeval}. The Tacotron2 model had the lowest naturalness scores, while %worst performance in terms of naturalness, and
Meta-IPA performed better than Meta-char. 
Our proposed model achieved the highest naturalness among the four models for all five languages. 
The MOS differences between our proposed model and baseline models were significant according to the confidence intervals, except the difference between our model and Meta-IPA on Dutch.
%Especially in Mandarin, German, French, Russian, our model significantly outperforms other models (p-value less than 0.05). Using our method will close the MOS gap between the synthesised speech and the recording. 
This confirms the effectiveness of our proposed method on improving the naturalness of multilingual speech synthesis.

\subsection{Encoder analysis}

% \begin{figure}[t]
% \centering
% \includegraphics[width=\linewidth]{frencoder.png}
% \caption{t-SNE visualization ofstress language (French) encoders. a is the output of the pronunciation encoder, b is the output of the prosody encoder, and c is the encoder output of the meta-learning TTS. Red is the Primary stress phoneme, yellow is the Secondary stress phoneme, the rest of the vowels are represented in purple, blue is the voiced consonant, and black is the unvoiced consonant. }
% \label{fig:fr_enocder}
% \end{figure}
\begin{figure}[t!]
\centering
 \includegraphics[width=\linewidth]{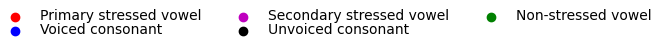}
\begin{subfigure}[t]{0.32\linewidth}
 \centering
 \includegraphics[width=\linewidth]{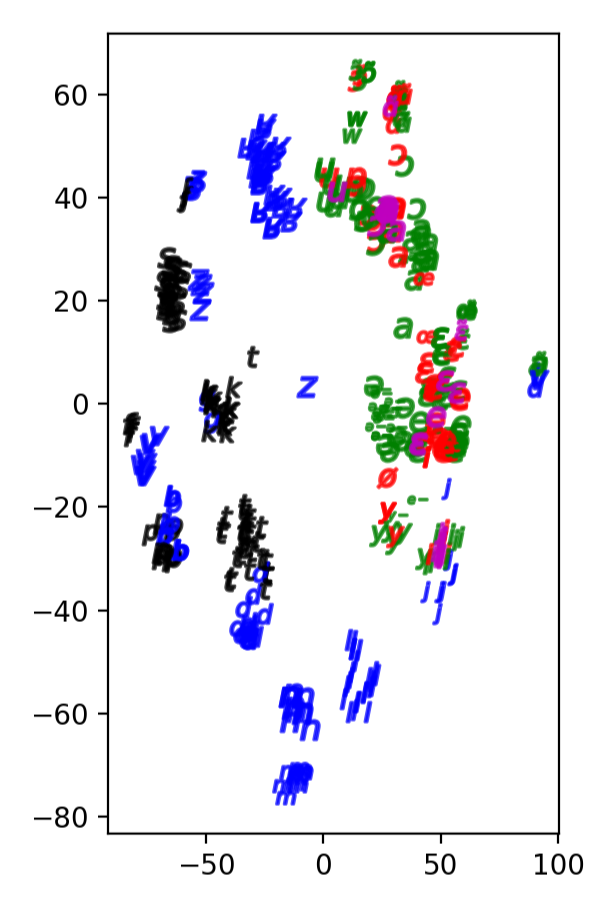}
\caption{}
\label{fig:a}
\end{subfigure}
\begin{subfigure}[t]{0.32\linewidth}
\centering
\includegraphics[width=\linewidth]{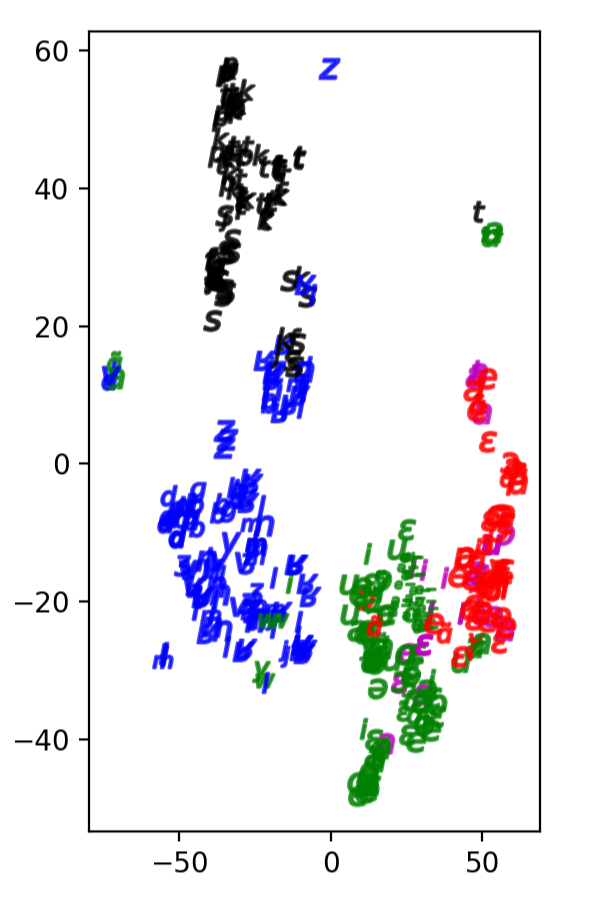}
\caption{}
\label{fig:b}
\end{subfigure}
\begin{subfigure}[t]{0.325\linewidth}
\centering
\includegraphics[width=\linewidth]{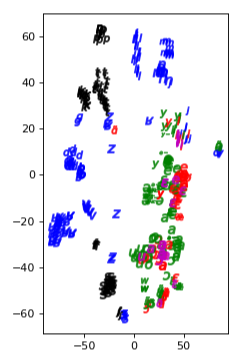}
\caption{}
\label{fig:c}
\end{subfigure}
% \begin{subfigure}[t]{0.3\linewidth}
% \centering
% \includegraphics[width=\linewidth]{fr2.png}
% \caption{}
% \label{fig:c}
% \end{subfigure}
% \begin{subfigure}[t]{0.15\textwidth}
% \centering
% \includegraphics[width=\textwidth]{zh1e1d.png}
% \caption{}
% \label{fig:c}
% \end{subfigure}
% \begin{subfigure}[t]{0.3\textwidth}
% \centering
% \includegraphics[width=\textwidth]{zh.png}
% \caption{}
% \label{fig:d}
% \end{subfigure}
\vspace{-3mm}
\caption{t-SNE visualization of the outputs of (a) the pronunciation encoder of the proposed model, (b) the prosody encoder of the proposed model, and (c) the encoder of Meta-IPA for French. Each point corresponds to a phoneme with its phonetic symbol. }
\label{fig:fr_enocder}
\vspace{-6mm}
\end{figure}

To verify whether our model can decouple pronunciation and prosody representations effectively, the encoder outputs calculated from the test utterances of French were visualized by the t-SNE algorithm, as shown in Fig.~\ref{fig:fr_enocder}.
Different colors discriminate the stress categories. For better visualization, we divided the category of non-stressed phoneme into three sub-categories, i.e., non-stressed vowel, voiced consonant and unvoiced consonant.

% Considering the large amount of non-stressed vowels, the phonemes of this category were ignored when drawing these figures.

We can see that the outputs of the pronunciation encoder in our model had similar distribution to that of the encoder in Meta-IPA, %have high similarity with their meta-learning TTS counterparts, 
i.e., the phonemes with the same phonetic symbol were close to each other. 
In Fig.~\ref{fig:fr_enocder}~(a), it can be further noticed that the instances of the consonants that have the same place and manner of articulation but only differ on voicing, e.g. /t/ and /d/, were also close to each other.
On the other hand, French is a non-tonal language with stresses on vowels. In Fig.~\ref{fig:fr_enocder}~(b), it can be observed that the outputs of the prosody encoder in our model were clustered according to phoneme categories, but secondary stressed vowel are difficult to distinguish from primary stressed vowel.
%In addition, we noticed that the proposed method has more obvious paired consonant, We think this may explain the reduction in V/UV-ERR. On the other hand, we also analyzed Chinese. In prosody encoder, all phonemes are divided into four categories according to their properties, including 
Similar patterns can also be observed in the outputs of the prosody encoder for Mandarin with tone categories which can't be shown in this paper due to limited space.
%This indicates that our model has the ability to learn phoneme pronunciation and prosody separately. Different from accent languages, vowels in tonal languages have different tones, and the presence of tones will cause the fundamental frequency of vowels to change during the pronunciation process. %show ~\ref{fig:zh_enocder}, 
%In prosody encoder, the phonemes are divided into four categories according to the tone. This shows that our model can adaptively learn prosody knowledge for each language.
All these results show that the outputs of the pronunciation encoder and the prosody encoder in our proposed model can capture the pronunciation and prosody characteristics of different languages separately, which indicates the effectiveness of decoupled modeling.

% \begin{figure}[t!]
% \centering
% \begin{subfigure}[t]{0.3\linewidth}
% \centering
% \includegraphics[width=\linewidth]{zh1e1d.png}
% \caption{}
% \label{fig:a}
% \end{subfigure}
% \begin{subfigure}[t]{0.6\linewidth}
% \centering
% \includegraphics[width=\linewidth]{zh.png}
% \caption{}
% \label{fig:b}
% \end{subfigure}
% \caption{t-SNE visualization oftone language (Chinese) encoders’ Vowel phonemes. a is the output of the pronunciation encoder, b is the output of the prosody encoder, and c is the encoder output of the meta-learning TTS. Blue, green, purple and red are the first, second, third and fourth tones respectively. }
% \label{fig:zh_enocder}
% \end{figure}

% \subsection{Discussion}

% In order to exclude the influence of parameters and output features on the results, based on Mate-IPA, we did a control experiment of doubling the parameters and using 80-dimensional Mel cepstrum as output features. The results showed that these changes did not significantly affect the pronunciation and prosody of speech. 

\section{Conclusions}

This paper proposed a two-stream model structure under the meta-learning framework to achieve decoupled pronunciation and prosody modeling for multilingual TTS. 
%method to improve speech quality by decoupling prosody and pronunciation features.
IPA symbols and prosody labels are employed as model input, and spectral features and excitation features are used as the prediction targets of the two streams, respectively.
Experimental results have shown that our proposed model significantly outperformed the baseline models in both objective and subjective evaluations.

\bibliographystyle{IEEEtran}

\bibliography{mybib}

\end{document}